\documentclass[letterpaper,fleqn,usenatbib]{mnras}

\usepackage{mathptmx}

\usepackage[T1]{fontenc}
\usepackage{ae,aecompl}

\usepackage{graphicx}	
\usepackage{amsmath}	
\usepackage{amssymb}	
\usepackage{xcolor}
\usepackage{booktabs}
\usepackage{relsize}
\usepackage{setspace}

\defcitealias{FKP}{FKP}
\defcitealias{PVP}{PVP}

\title[Optimal Multi-tracer Weights]{Optimal Weights For Measuring Redshift Space Distortions 
in Multi-tracer Galaxy Catalogues}

\author[D. W. Pearson, L. Samushia, and P. Gagrani]{
David W. Pearson$^{1}$\thanks{E-mail: dpearson@phys.ksu.edu},
Lado Samushia$^{1,2,3}$,
and Praful Gagrani$^{1}$
\\
$^{1}$Department of Physics, Kansas State University, 116 Cardwell Hall, Manhattan, KS, 66506, 
USA\\
$^{2}$National Abastumani Astrophysical Observatory, Ilia State University, 2A Kazbegi Ave., 
GE-1060 Tbilisi, Georgia\\
$^{3}$Institute of Cosmology \& Gravitation, University of Portsmouth, PO1 3FX, UK
}

\date{Accepted 2016 August 25. Received 2016 August 24; in original form 2016 June 10}

\pubyear{2016}

\begin{document}
\label{firstpage}
\pagerange{\pageref{firstpage}--\pageref{lastpage}}
\maketitle

\begin{abstract}
Since the volume accessible to galaxy surveys is fundamentally limited, it is
extremely important to analyse available data in the most optimal fashion. One
way of enhancing the cosmological information extracted from the clustering of
galaxies is by weighting the galaxy field. The most widely used weighting
schemes assign weights to galaxies based on the average local density in the
region (\textit{FKP weights}) and their bias with respect to the dark matter
field (\textit{PVP weights}). They are designed to minimize the fractional
variance of the galaxy power-spectrum. We demonstrate that the currently used bias
dependent weighting scheme can be further optimized for specific cosmological
parameters. We develop a procedure for computing the optimal weights and test
them against mock catalogues for which the values of all fitting parameters, as
well as the input power-spectrum are known. We show that by applying these
weights to the joint power-spectrum of Emission Line Galaxies and Luminous Red
Galaxies from the Dark Energy Spectroscopic Instrument survey, the variance in
the measured growth rate parameter can be reduced by as much as 36 per cent.
\end{abstract}

\begin{keywords}
methods: data analysis -- large-scale structure of the Universe -- galaxies: statistics --
cosmological parameters
\end{keywords}



\section{Introduction}

Future galaxy redshift surveys, such as the Dark Energy Spectroscopic Instrument
survey \citep[DESI;][]{Schlegel2011,Levi2013}, the Extended Baryon Oscillation
Spectroscopic survey \citep[eBOSS;][]{Schlegel2009}, the \textit{Euclid}
satellite surveys \citep{Laureijs2011}, and the Wide Field Infrared Survey 
Telescope surveys \citep[WFIRST;][]{Spergel2013} will cover vast cosmological volumes 
with a high number density of galaxies. Since the available cosmic volume is 
fundamentally limited a lot of effort is going into developing optimal ways of 
analysing galaxy clustering data \citep[see e.g.,][]
{Eisenstein2007,Blazek2014,Bianchi2015,Scoccimarro2015,Slepian2016}.

One way of improving the variance of measured 2-point statistics is to weight
the galaxy field to achieve the optimal signal-to-noise. The most commonly used
weighting scheme is the one developed by \citet*[hereafter
\citetalias{FKP}]{FKP}, which is used in all analyses employing 2-point
statistics \citep[see e.g.,][]{Percival2001,Reid2010,Blake2011,Anderson2012,
Ross2013,Beutler2011,Beutler2012,Beutler2014,Gil-Marin2015}. The FKP weights, 
\begin{equation} 
\label{eq:wfkp} w_{\mathrm{FKP}}(\bmath{r}) \propto
\dfrac{1}{1+\overline{n}(\bmath{r})P(\bmath{k})}, 
\end{equation} 
where $\overline{n}(\bmath{r})$ is the average number density of galaxies at a position
$\bmath{r}$ and $P(\bmath{k})$ is the power-spectrum at a wavelength of interest
$\bmath{k}$, are straightforward to apply and reduce the variance of the
measured power-spectrum when the completeness of the galaxy sample is
significantly non-uniform. 

\citet*[hereafter \citetalias{PVP}]{PVP} further optimized the FKP scheme for
samples that include galaxies with a range of biases with respect to the dark
matter. If the number density is uniform the PVP weights are 
\begin{equation} 
\label{eq:wfkp} w_{\mathrm{PVP}} \propto b,
\end{equation} 
where $b$ is the bias with respect to the dark matter, and will minimize the fractional 
variance in the measured power-spectrum.

If a galaxy sample covers such a wide redshift range that the effects of cosmic
evolution are significant, the measured power-spectrum will be a weighted average
of power spectra at different redshifts within that range. Since the sensitivity
of the power-spectrum to cosmological parameters also varies with redshift, it
is possible to construct redshift dependent weights which maximize the
constraining power of the measured power-spectrum for specific cosmological parameters.
This optimal weighting scheme obviously depends on which cosmological parameters
we want to optimize. Recently, \citet*{Zhu2015} derived redshift weights that
optimize the measurement of the Baryon Acoustic Oscillation (BAO) Peak position,
while \citet{Ruggeri2016} derived similar weights that optimize the Redshift
Space Distortion (RSD) parameter. Both works built on the formalism developed in
\citet*{Tegmark1997}.

Future surveys will observe emission line galaxies (ELGs), luminous red galaxies
(LRGs), and quasars (QSOs), in overlapping volumes. Computing power spectra of
all tracers individually is suboptimal as important information encoded in the
cross-correlation of the tracers will be lost. A promising way of taking
advantage of the presence of multiple tracers was proposed in
\citet{McDonald2009}. The method, however, has not yet been implemented in
practice and will only result in a significant improvement in constraining power
when the number density of tracers is high (which is not the case e.g. in the
eBOSS survey).  Computing all auto and cross power spectra is possible \citep{Ross2012}
but will require accurate estimation of large
covariance matrices which may be problematic for future surveys
\citep{Pope2008,Schneider2011,dePutter2012,Xu2012,delaTorre2013,Mohammed2014,Paz2015,
Grieb2016,Pearson2016}. The most straightforward way is to
compute a single power-spectrum for all tracers while differentially weighting
them to achieve optimal signal-to-noise by applying the PVP weights.

While the PVP weights are designed to minimize the fractional variance in the 
power-spectrum, this does not necessarily translate into minimal variance on measured
cosmological parameters. A good example of this is the growth rate parameter,
$f$.\footnote{In practice, from the galaxy clustering data alone the growth rate
parameter is measured up to a normalization constant $f\sigma_8$. We will be
using $f$ to mean the $f\sigma_8$ combination for brevity.  This has no effect on
our results.} The growth rate is measured from an anisotropic signature in the
power-spectrum which is more pronounced for low biased tracers.  The power-spectrum 
signal on the other hand is stronger in the high biased tracers. The
weights in equation~\eqref{eq:wfkp} upweight high bias galaxies to achieve the
optimal power-spectrum signal, but the measured power-spectrum becomes less
sensitive to $f$. The optimal weighting for the growth rate parameter must
counterbalance these two tendencies by producing a power-spectrum with a small
(not necessarily minimal) variance that is at the same time sensitive enough to
the $f$ parameter.

In this work we generalize the \citetalias{PVP} weighting scheme to minimize the 
variance of specific cosmological parameters measured from the power-spectrum
(section~\ref{sec:weighting}). We assume that the galaxy samples will be analysed
in narrow redshift bins of $\delta z \sim 0.1$, eliminating the need to
consider the redshift evolution weights \citep{Zhu2015,Ruggeri2016} as the
effect will be small. We test our new
weighting scheme on mock catalogues of the eBOSS and DESI surveys
(section~\ref{sec:analysis}) and show that they could improve the variance of the
measured $f$ parameter by up to 36 per cent.  As expected, the optimal
weights differ for different cosmological parameters (section~\ref{sec:results}).
This weighting scheme is straightforward to compute and implement and should
result in reduced variance on cosmological parameters measured from future
galaxy surveys (section~\ref{sec:conclusion}).

\section{Optimal Weighting}
\label{sec:weighting}
For simplicity, we will assume that galaxies of two types with densities
$n_1(\bmath{r})$ and $n_2(\bmath{r})$ are present in an overlapping volume and
the average number densities $\overline{n}_1$ and $\overline{n}_2$ do not vary
significantly within the volume. The formalism is easy to generalize for more
than two tracers and varying number densities. If we assign weights $w_1$ and
$w_2$ to these galaxies, then the number density of the combined field is
\begin{equation}
n(\bmath{r})=w_1n_1(\bmath{r}) + w_2n_2(\bmath{r})
\end{equation}
and the overdensity field is
\begin{equation}
\label{eq:overdensity}
\delta(\bmath{r}) \equiv \frac{n(\bmath{r}) - \overline{n}}{\overline{n}} = 
A_1\delta_1(\bmath{r}) + A_2\delta_2(\bmath{r}),
\end{equation}
where the overdensities are defined by
\begin{equation}
\delta_i(\bmath{r}) = \frac{n_i(\bmath{r}) - \overline{n}_i}{\overline{n}_i},
\end{equation}
and
\begin{equation}
A_i = \frac{w_i\overline{n}_i}{w_1\overline{n}_1 + w_2\overline{n}_2}
\end{equation}
is the weighted fractional density. We will assume the weights to be normalised
by $w_1 + w_2 = 1$. This will shorten some of our formulas, although in practice
only the ratio of weights is relevant.
The power-spectrum of the overdensity field,
\begin{equation}
P(\bmath{k})\equiv\left|\widetilde{\delta}(\bmath{k})\right|^2,
\end{equation}
can be estimated from the squared modulus of the Fourier transform,
\begin{equation}
\widetilde{\delta}(\bmath{k}) =
\displaystyle\int\!\mathrm{d}\bmath{r}\,\mathrm{e}^{-i\bmath{kr}}\delta(\bmath{r}),
\end{equation}

We will assume that the overdensity fields are Gaussian (this is a common
assumption when deriving optimal weights) with
\begin{equation}
\label{eq:pk}
\left<\widetilde{\delta}_i(\bmath{k})\widetilde{\delta}^\star_j(\bmath{k})\right>=
\left[\left(b_i+\mu^2f\right)\left(b_j+\mu^2f\right)P_\mathrm{m}(k)
+ \frac{\delta^\text{\tiny C}_{ij}}{\overline{n}_i}\right]V_\mathrm{s},
\end{equation}
where the angular brackets denote the expectation value, $\delta^\text{\tiny C}$
is a Kronecker delta function, $V_\mathrm{s}$ is the survey volume, and
$P_\mathrm{m}(k)$ is the matter power-spectrum that can be computed in any given
cosmological model \citep{Kaiser1987,Hamilton1997}.
The last term in equation~\eqref{eq:pk} is the \emph{Shot-Noise term} due to the
sampling of the overdensity field with a finite number of galaxies 
\citepalias{FKP}.\footnote{The power-spectrum estimators are usually defined after 
subtracting the Shot Noise term, but this is irrelevant for our results.} For the 
weighted field in equation~\eqref{eq:overdensity} this results in 
\begin{equation}
\label{eq:pkmodel}
P(\bmath{k}) = \left[\left(b_\mathrm{eff}^{w_1w_2}+\mu^2f\right)^2P_\mathrm{m}(k) +
S^{w_1w_2}\right]V_\mathrm{s},
\end{equation}
with the weighting dependent effective bias,
\begin{equation}
b_\mathrm{eff}^{w_1w_2} =
A_1b_1
+
A_2b_2,
\end{equation}
and the shot noise term,
\begin{equation}
S^{w_1w_2} = 
\frac{A_1^2}{\overline{n}_1} + \frac{A_2^2}{\overline{n}_2}.
\end{equation}

Since we assumed the overdensity field to be Gaussian, the variance of the galaxy
power-spectrum estimator is simple to compute and is 
\begin{equation}
\mathrm{Var}\left[P(\bmath{k})\right] \propto
\left[\left(b_\mathrm{eff}^{w_1w_2}+\mu^2f\right)^2P_\mathrm{m}(k) +
S^{w_1w_2}\right]^2
\end{equation}
\citep[\citetalias{FKP}; \citetalias{PVP};][]{Tegmark1997}. The fractional variance in the
galaxy power-spectrum is then
\begin{equation}
\label{eq:fracvar}
\frac{P(\bmath{k})}{\mathrm{Var}\left[P(\bmath{k})\right]}\propto
\frac{\left(b_\mathrm{eff}^{w_1w_2}+\mu^2f\right)^2P_\mathrm{m}(k)}{\left[\left(b_\mathrm{eff}^{w_1w_2}+\mu^2f\right)^2P_\mathrm{m}(k)
+ S^{w_1w_2}\right]^2}.
\end{equation}
This expression is minimized\footnote{This can be verified by simply equating
the partial derivatives of equation~\eqref{eq:fracvar} with respect to the weights to
zero along with the Legandre multipliers to enforce the condition $w_1+w_2=1$.} by 
\begin{equation}
\label{eq:PVPweight}
\setstretch{2.2}
\begin{array}{@{}l}
w_1 = \dfrac{b_1}{b_1 + b_2},\\
w_2 = \dfrac{b_2}{b_1 + b_2},\\
\end{array}
\end{equation}
which, for constant number densities, is equivalent to \citetalias{PVP}
weighting.\footnote{For a more rigorous derivation also accounting for the number density
variations see \citetalias{PVP}.}

The minimum fractional variance in the power-spectrum, however, does not
necessarily correspond to the minimum variance in the cosmological parameters
derived from the power-spectrum. The power-spectrum is most sensitive to the
bias -- $b_{\mathrm{eff}}$, growth rate -- $f$, and the Alcock-Paczinsky parameters
$\alpha_{\parallel}$ and $\alpha_\perp$ \citep{Alcock1979,Kaiser1987}. The dependence on $b$ and $f$ is already in
equation~\eqref{eq:pkmodel}, and the dependence on $\alpha_\perp$ and
$\alpha_\parallel$ can be introduced by replacing
\begin{equation}
\label{eq:kAP}
k \longrightarrow \dfrac{k}{\alpha_{\perp}^{}}\left[1+\mu^{2}\left(\dfrac{\alpha_{\perp}^{2}}{
\alpha_{\parallel}^{2}} - 1\right)\right]^{1/2},
\end{equation}
\begin{equation}
\label{eq:muAP}
\mu \longrightarrow \dfrac{\mu\alpha_{\perp}^{}}{\alpha_{\parallel}^{}}\left[1+\mu^{2}\left(\dfrac{
\alpha_{\perp}^{2}}{\alpha_{\parallel}^{2}}-1\right)\right]^{-1/2},
\end{equation}
 and dividing the power-spectrum by a factor of $\alpha_\parallel\alpha_\bot^2$
\citep{Ballinger1996,Simpson2010,Samushia2011}.  The Fisher information matrix
of these parameters is
\begin{equation}
\label{eq:fishertot}
F_{ij}=\displaystyle\mathlarger{\int}\!\mathrm{d}\bmath{k}\,\frac{\partial P(\bmath{k})}{\partial
\theta_i}\frac{1}{\mathrm{Var}\left[P(\bmath{k})\right]}\frac{\partial P(\bmath{k})}{\partial
\theta_j},
\end{equation}
where $\bmath{\theta}=(b_{\mathrm{eff}},f,\alpha_\parallel,\alpha_\bot)$ is a parameter vector,
and the integration is over all wavevectors, the power-spectrum measurements of
which where used in the analysis. The inverse of the Fisher matrix gives a
covariance matrix
\begin{equation}
\label{eq:covtot}
\mathsf{C}=\mathsf{F}^{-1}
\end{equation}
and the diagonal elements of the covariance matrix correspond to the expected
variance of the parameters measured from the power-spectrum. Because of the
presence of the derivative terms (that also depend on the weights) in
equation~\eqref{eq:fishertot} the weighting scheme that minimizes the variance of the
power-spectrum does not necessarily minimize the diagonal elements of the
covariance matrix in equation~\eqref{eq:covtot}. 

A simple analytic solution for the optimal weights in this case does not exist,
but they are relatively straightforward to find numerically.  To find the
optimal weights we numerically compute the variance and the derivatives in
equation~\eqref{eq:fishertot} and take the integral over the wavevectors of interest.
We then numerically find the weights that minimize the diagonal elements of the
inverse Fisher matrix of equation~\eqref{eq:covtot}.\footnote{Since we adopt the
normalization $w_1 + w_2 = 1$ this turns into a simple one parameter
minimization procedure.} These weights will in general be different for each
parameter. In the tradition of \citetalias{FKP} and \citetalias{PVP}, we refer
to these as PSG weights in what follows.

\section{Data and Measurement Procedures}
\label{sec:analysis}

\subsection{The Sample}
\label{sec:themocks}
In order to test our weighting scheme, we generated lognormal mock catalogues
\citep{Coles1991} giving us control over the input power-spectrum and linear
growth rate. 

We computed the matter power-spectrum using the \textsc{camb} software
\citep*{Lewis2000} via the web interface hosted at
LAMBDA\footnote{\href{http://lambda.gsfc.nasa.gov/toolbox/tb_camb_form.cfm}
{http://lambda.gsfc.nasa.gov/toolbox/tb\_camb\_form.cfm}} for a spatially flat
$\Lambda$CDM cosmology with $\Omega_{\mathrm{M}}=0.276$, and $\Omega_{\mathrm{b}}h^2=0.0226$. 
We use the fiducial value of 
\begin{equation}
\label{eq:growth}
f(\Omega_{\mathrm{M}},z) \approx \Omega_{\mathrm{M}}^{0.6}(z),
\end{equation}
where
\begin{equation}
\label{eq:matterden}
\Omega_{\mathrm{M}}(z) = \dfrac{\Omega_{\mathrm{M},0}(1+z)^{3}}{\Omega_{\mathrm{M},0}(1+z)^{3}
+\Omega_{\Lambda,0}},
\end{equation}
which is the value predicted by general relativity
\citep{Peebles1980,Martinez2002}.

Our lognormal code was largely based on the description given in
\citet{Weinberg1992} and Appendix A of \citet{Beutler2011}, with modifications
required to obtain a distribution of two tracers cross-correlated by the same
underlying matter field.

We started by distributing the power to two grids -- one for LRGs and one for ELGs --
in $k$-space as 
\begin{equation}
\label{eq:kaiser} P(\bmath{k}) = \left(b_i + \mu^{2}f\right)^{2}P_{\mathrm{m}}(k),
\end{equation} 
where $\mu = k_z/k$ and $b_i$ is the bias of LRGs or ELGs for the redshift bin. 
This was assigned to the real part only, with the imaginary part being set to zero. 
After performing inverse Fourier transforms using the complex-to-real transform 
in the Fastest Fourier Transform in the West (\textsc{fftw}) 
library\footnote{\href{http://fftw.org/}{http://fftw.org/}} \citep{Frigo2005}, 
we took the resulting correlation functions and calculated
$\ln[1+\xi^{i}(\bmath{r})]$ at each grid point, then performed real-to-complex transforms.
The result of these transforms, $P^{i}_{\mathrm{LN}}(\bmath{k})$, was then normalized by the 
number of grid points since \textsc{fftw} produces the unnormalized Fourier transform.

At this stage, we took the ratio of the $P^{i}_{\mathrm{LN}}(\bmath{k})$ at each grid point
in $k$-space and stored that in memory. We then constructed Gaussian random
realizations by drawing from a normal distribution, centred on zero, with 
$\sigma = \sqrt{\max\lbrace0,
\mathrm{Re}[P_{\mathrm{LN}}(\bmath{k})]\rbrace/2}$, at each grid
point for both the real and
imaginary parts, in order to obtain a well behaved power-spectrum \citep{Weinberg1992}.
We took care that $\delta_{\mathrm{LN}}(-\bmath{k}) =
\delta_{\mathrm{LN}}^{\ast} (\bmath{k})$, and that the values at grid points
whose indices are combinations of zero and $N_{i}/2$, where $N_{i}$ is the
number of grid points in dimension $i$, were purely real. This ensured that when
we inverse Fourier transformed $\delta_{\mathrm{LN}}(\bmath{k})$, the result was
purely real. \emph{We only did the random draw for the higher bias tracer,
then using the ratio previously calculated, we scaled the random realization to
obtain the values on the grid for the lower bias tracer}. In this way, we were
able to effectively obtain mock samples with two tracers, each following the
same underlying matter distribution.

The last step was then to take the inverse Fourier transforms of the random
realization for the higher bias tracer, and the scaled realization for the lower
bias tracer. This resulted in overdensity fields for both tracers,
$\delta_{i}(\bmath{r})$, having zero mean and variance
$\sigma_{\mathrm{G}}^{2}$. From these overdensity fields we calculated the
lognormal density field
\begin{equation}
\delta_{\mathrm{LN},i}(\bmath{r}) = \exp\left[\delta_{i}(\bmath{r})-\sigma_{\mathrm{G}}^{2}/2
\right].
\end{equation}
This was then multiplied by the average number of galaxies per cell to give the
desired number density, and Poisson sampled to create our final galaxy
catalogues, placing the galaxies randomly within a given cell.

\begin{table*}
\centering
\caption{Mock catalogue properties. Column 1 list the redshift range. Columns 2, 3
and 4 list the volume of the cube and the number denisties for LRGs and ELGs to match the eBOSS
survey, respectively. Columns 5, 6, and 7 list the volume of the cube and the number densities
for LRGs and ELGs to match that expected in the DESI survey, respectively. Columns 8 and 9 list
the biases for the LRGs and ELGs, respectively. Lastly, column 7 lists the dimensionless linear
growth rate for the redshift bin.}
\label{tab:mockprops}
\begin{tabular}{@{}lccccccccc@{}}
\toprule
Redshift & $V_{\mathrm{eBOSS}}$ & $\bar{n}_{\mathrm{LRG,eBOSS}}$ & 
$\bar{n}_{\mathrm{ELG,eBOSS}}$ & $V_{\mathrm{DESI}}$ & $\bar{n}_{\mathrm{LRG,DESI}}$ & 
$\bar{n}_{\mathrm{ELG,DESI}}$ & $b_{\mathrm{LRG}}$ & $b_{\mathrm{ELG}}$ & $f$ \\
& ($h^{-3}\mathrm{Gpc}^{3}$) & $(10^{-4}h^{3}\mathrm{Mpc^{-3}})$ & 
$(10^{-4}h^{3}\mathrm{Mpc^{-3}})$ & ($h^{-3}\mathrm{Gpc}^{3}$) & 
$(10^{-4}h^{3}\mathrm{Mpc^{-3}})$ & $(10^{-4}h^{3}\mathrm{Mpc^{-3}})$\\
\midrule
$0.6 \leq z < 0.7$ & 0.272 & 0.772 & 1.345 & 2.538 & 4.589 & 1.704 & 2.339 & 1.376 & 0.759 \\
$0.7 \leq z < 0.8$ & 0.325 & 0.642 & 2.051 & 3.031 & 4.555 & 10.482 & 2.450 & 1.441 & 0.787 \\
$0.8 \leq z < 0.9$ & 0.374 & 0.330 & 1.559 & 3.491 & 2.655 & 7.711 & 2.563 & 1.508 & 0.812 \\
$0.9 \leq z < 1.0$ & 0.419 & 0.091 & 0.586 & 3.914 & 0.973 & 7.490 & 2.678 & 1.575 & 0.834 \\
\bottomrule
\end{tabular}
\end{table*}

We generated 1024 mock catalogues for four redshift bins in $0.6<z<1.0$. They
contained two tracers designed to mimic LRGs and ELGs. This led to a total of
4096 mock catalogues with number densities to match those expected in the eBOSS
survey, and 4096 with number densities to match those expected in the DESI survey.
Table~\ref{tab:mockprops} lists the specific properties for the mock catalogues.
For the eBOSS mocks, the number densities were calculated from information in
\citet{Zhao2016}. The volumes were calculated for the 1500~deg$^2$ region were
the ELGs and LRGs would overlap. The DESI number densities were calculated from
information in the DESI Science Final Design Report \citep{DESISciTDR}, and the
volumes assume the 14,000 deg$^2$ baseline survey footprint. The biases for the
different redshift bins were given by \citet{Dawson2016}.

To have a clean separation between number density dependent and bias dependent
weights, our mock catalogues have a constant number density. Since there
are no number density gradients, the \citetalias{FKP} weights reduce to a simple uniform
weighting (see section~\ref{sec:results}), meaning any improvements are only coming from the
differential weighting of tracers based on their bias.

\subsection{Measuring the Power-Spectrum}
\label{sec:measpk}
We followed the general methods of \citetalias{FKP} for measuring the power-spectrum
from our mock catalogues. We generated random catalogues with 10 or 30
times the number of each tracer for the DESI and eBOSS mocks, respectively. The
galaxies and randoms were binned using cloud-in-cell interpolation \citep{Birdsall1969} 
with one of the three different weighting schemes
-- see section~\ref{sec:results} for details.  Since we used a discrete Fourier
transform of boxes with a finite linear size $L = V^{1/3}$ (see Table \ref{tab:mockprops}
for volumes), our
$\tilde{\delta}(\bmath{k})$ (and correspondingly $P(\bmath{k})$ measurements) were
given on a discrete cubic grid with a resolution of $2\pi/L$. To compress this
information we computed the spherically averaged power-spectrum monopole and
quadrupole ($l = 0$, 2) in 24 bins of width
$\Delta k = 0.008$ for $0.008 \leq k \leq 0.2$ via
\begin{equation}
\label{eq:pkbin}
P_{l}(k) = \dfrac{2l+1}{2}\sum 
\left|\widetilde{\delta}(\bmath{k})\right|^{2}\mathcal{P}_{l}\left[\mu(\bmath{k})\right]G^{2}(\bmath{k}),
\end{equation}
where the sum is over all wavevectors in the range $k-\Delta k/2 \leq
|\bmath{k}| < k+\Delta k/2$, $\mathcal{P}_{l}(x)$ are the Legendre polynomials,
and $G(\bmath{k})$ is a grid correction term
\begin{equation}
\label{eq:windowcor}
G(\bmath{k}) = \prod_{i} \left[\mathrm{sinc}(\Delta L_{i}k_{i})\right]^{-2},
\end{equation}
with $\mathrm{sinc}(x) = \sin(x)/x$, $\Delta L_{i} = L_{i}/N_{i}$, $i$ denotes
one of the three Cartesian coordinates, $L_{i}$ is the length of the cube in
that coordinate direction, and $N_{i}$ is the corresponding number of grid points. 

We only considered the power-spectrum measurements below $k\sim0.2$ allowing us to
safely ignore the non-linear effects at 
higher wavevectors which are difficult to model and are usually excluded from the
analysis. Even though our Fisher matrix predictions in section \ref{sec:weighting} implicitly
assumed that each $P(\bmath{k})$ measurement would be analysed individually
without reducing them to the multipoles, we do not expect this to be a big
effect as a number of recent studies showed that reducing the power-spectrum to
the first few even multipoles retains most of the information \citep{Taruya2011,Kazin2012}.

The measurements resulting from equation~\eqref{eq:pkbin} were then corrected for shot
noise
\begin{equation}
\label{eq:shotnoise}
S_{l}(k) = \dfrac{1}{N_{k}}\left(\sum_{\mathrm{gal}}w_{i}^{2} + \alpha^{2}\sum_{\mathrm{ran}}
w_{i}^{2}\right)\sum_{\mu}\mathcal{P}_{l}(\mu),
\end{equation}
where the sum is over the same modes, with $\alpha =
\sum_{\mathrm{gal}}w_{i}/\sum_{\mathrm{ran}}w_{i}$, and then normalized by
\begin{equation}
\label{eq:normalization}
I =
V_\mathrm{s}\left(\overline{n}_{\mathrm{LRG}}w_{\mathrm{LRG}}+\overline{n}_{\mathrm{ELG}}w_{\mathrm{ELG}}\right)^{2}.
\end{equation}

Since the power-spectrum multipoles in equation~\eqref{eq:pkbin} were computed as a
discrete sum over a finite number of wavevectors, modelling them as angular
integrals over the theoretical power-spectrum of equation~\eqref{eq:kaiser} would be
inaccurate. To model the $\bmath{k}$-grid discreteness effects we 
distributed the model power to the same grid
used to calculate the power-spectrum from the mocks and binned it according to 
equation~\eqref{eq:pkbin} to give $P^{\mathrm{grid}}_{l}(k)$. We then adjusted the
measured value,
\begin{equation}
\label{eq:corrPk}
P_{l}(k) \longrightarrow P_{l}(k)-\left[P^{\mathrm{grid}}_{l}(k)-P^{\mathrm{int}}_{l}(k)\right],
\end{equation}
where 
\begin{equation}
\label{eq:exactPk}
P^\mathrm{int}_{l}(k) = \dfrac{2l+1}{2}\displaystyle\int\limits_{-1}^{1}\!\left(b_{\mathrm{eff}}^{w_{1}w_{2}}
+\mu^{2}f\right)^{2}P_{\mathrm{m}}(k)\mathcal{P}_{l}(\mu)~\mathrm{d}\mu,
\end{equation}
is the integrated power-spectrum \citep{Blake2011,Beutler2014}. This effect
(correction term in the brackets) is extremely small for the monopole, so in
practice we only applied the correction to the quadrupole.

\begin{figure}
\includegraphics[width=1.0\linewidth]{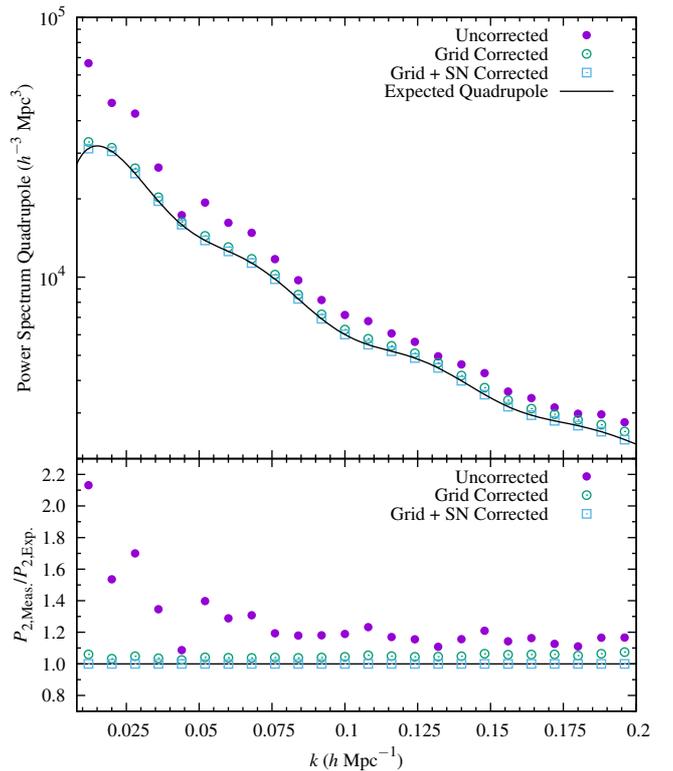}
\caption{Here we show the effects of the discreteness of the grid on the
quadrupole for the number densities and volume of the first redshift bin of
eBOSS. The solid circles (purple) show the model power after being distributed
to the grid, binned as in equation~\eqref{eq:pkbin} and adding the expected
shot noise calculated as in equation~\eqref{eq:shotnoise}. The open
circles (green) show the effects of the correction in
equation~\eqref{eq:corrPk}. It is clear that even after applying that
correction, there is a small positive bias (${\sim}5$ per cent). The open
squares (light blue) show the measured quadrupole after correcting for the
discreteness in the shot-noise as well, at which point we have recovered the
expected quadrupole quite well.}
\label{fig:pkcorr}
\end{figure}

We would like to emphasize that, given the small volume of our mocks (especially
the eBOSS mocks), the $\bmath{k}$-grid discreteness effects also had to be
accounted for when computing the shot-noise correction. Even though integrals over
higher order Legendre polynomials are zero, the discrete sum over $\mu$ in
equation~\eqref{eq:shotnoise} is nonzero. This implies that the shot-noise
corrections have to be applied not only to the monopole but to the higher order
multipoles of the power-spectrum as well. Fig.~\ref{fig:pkcorr} explicitly shows
these effects on the quadrupole for the first redshift bin eBOSS mocks. We found
that ignoring the effects in equation~\eqref{eq:corrPk} can bias the quadrupole
by ${\sim}27$ per cent on average, and significantly more so at low wavenumbers.
While significantly smaller, ignoring the discreteness effects in the shot-noise
--~equation~\eqref{eq:shotnoise}~-- can bias the quadrupole by ${\sim}5$ per
cent. We note that the size of these corrections decrease for increased volume,
and the shot-noise additionally decreases for increased number densities.  For
example, in the first redshift bin of the DESI mocks, the effect in
equation~\eqref{eq:corrPk} drops to ${\sim}13$ per cent on average, while the
shot-noise causes a bias of less than 1 per cent.

\begin{figure*}
\includegraphics[width=1.0\linewidth]{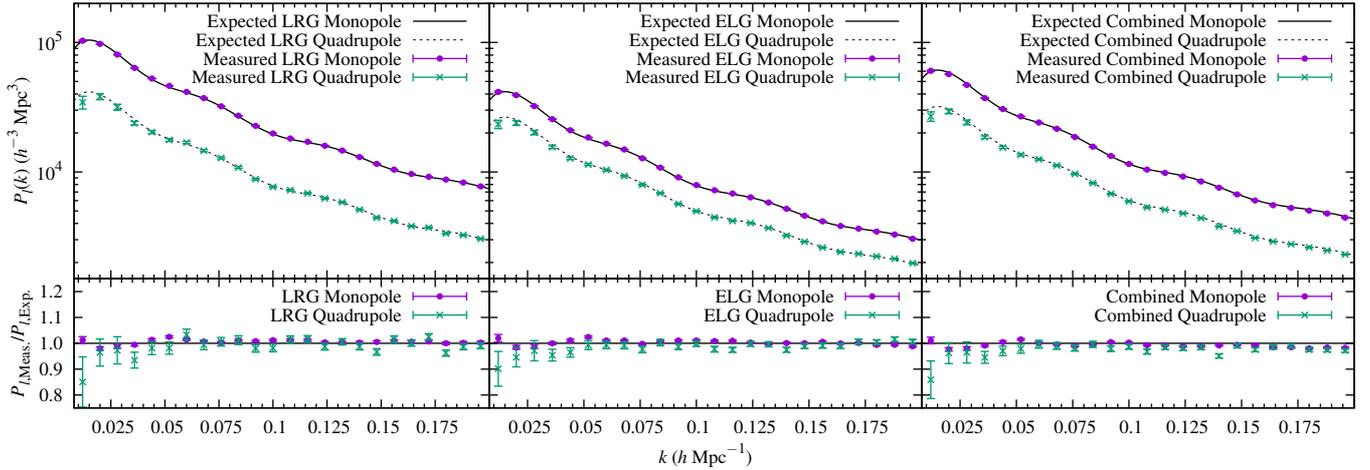}
\caption{Comparison of the expected power-spectrum monopole and quadrupole with
the measured values after applying all corrections for LRGs (left), ELGs
(middle) and the combined sample (right) with equal weighting in the first
redshift bin with eBOSS number densities. The plotted error bars are
$\sqrt{C_{ii}/N}$ where $C_{ii}$ are the diagonal elements of the sample
covariance matrix, and $N$ is the number of mocks.}
\label{fig:mockverify}
\end{figure*}

Fig. \ref{fig:mockverify} shows a detailed comparison of the power-spectrum
which we expected to recover and the power-spectrum that we actually measured
for the first redshift bin ($0.6 \leq z < 0.7$) with the eBOSS number densities
and volume. We explicitly show the recovered power for each tracer individually
as confirmation of our scaling procedure (see section~\ref{sec:themocks}). We
also note that we recovered the expected power-spectrum extremely well for the other
eBOSS redshift bins and for the DESI volumes as well. 


\subsection{Parameter Estimation}
\label{sec:paraest}
We model the measured power-spectrum multipoles as
\begin{equation}
\label{eq:modelPk}
P_{l}(k) =
\dfrac{2l+1}{2\alpha_{\perp}^{2}\alpha_{\parallel}^{}}\displaystyle\int\limits_{-1}^{1}
\!\mathrm{d}\mu~P(k,\mu)\mathcal{P}_{l}(\mu),
\end{equation}
where
\begin{equation}
\label{eq:galPk}
P(k,\mu) = \left(b_{\mathrm{eff}}^{w_{1}w_{2}} +
\mu^2f\right)^{2}P_{\mathrm{m}}(k),
\end{equation}
and $k$, and $\mu$ depend on $\alpha_\perp$ and $\alpha_\parallel$ as in
equations~\eqref{eq:kAP}~and~\eqref{eq:muAP}. The shape of the power-spectrum is
fixed, while the four parameters $b_{\mathrm{eff}}^{w_{1}w_{2}}$, $f$,
$\alpha_\parallel$, $\alpha_\perp$ are free.\footnote{Hereafter, we simply refer to $b_{\mathrm{eff}}^{w_{1}w_{2}}$ as $b$ for simplicity.}

In order to find the best-fitting parameters, we used a Markov-Chain Monte Carlo
(MCMC) method with the Metropolis-Hastings algorithm \citep{Hastings1970} to
find the posterior likelihood of the free parameters. In all our MCMC chains the
mean values of the parameters were very close to the input values and the
likelihood surfaces were close to Gaussian. We computed the variance of each
parameter from the MCMC mocks and compared the resulting values for all the
parameters for a specific set of weights to other weighting schemes to see if
the PSG weights actually yield the tightest constraints.

\section{Comparison with Other Weighting Schemes}
\label{sec:results}

In order to test the weights purely from the stand point of relative weighting
of tracers, our mock catalogues have uniform number densities through out, and
only two types of tracers with constant biases. In what follows, for brevity we
will quote weights as pairs in the form $(w_{\mathrm{LRG}}, w_{\mathrm{ELG}})$. 

\begin{table*}
\caption{The PSG weights for the free parameters in our model. Column 1
indicates the target survey for the weights. Column 2 gives the redshift
range. Columns 3, 4, 5, and 6 give the optimal weights for each parameter
individually.}
\label{tab:weights}
\centering
\begin{tabular}{@{}lccccc@{}}
\toprule
Survey & Redshift Range & $b$ & $f$ & $\alpha_{\perp}$ & $\alpha_{\parallel}$ \\
\midrule
DESI & $0.6 \leq z < 0.7$ & $(0.279,0.721)$ & $(0.207,0.793)$ & $(0.546,0.454)$ & 
                                                      $(0.464,0.536)$ \\
     & $0.7 \leq z < 0.8$ & $(0,1)$ & $(0,1)$ & $(0.516,0.484)$ & 
                                                      $(0.347,0.653)$ \\
     & $0.8 \leq z < 0.9$ & $(0,1)$ & $(0,1)$ & $(0.547,0.453)$ & 
                                                      $(0.429,0.571)$ \\
     & $0.9 \leq z < 1.0$ & $(0,1)$ & $(0,1)$ & $(0.568,0.432)$ & 
                                                      $(0.472,0.528)$ \\[1ex]
eBOSS& $0.6 \leq z < 0.7$ & $(0.427,0.573)$ & $(0.367,0.633)$ & $(0.589,0.411)$ & 
													  $(0.547,0.453)$ \\
     & $0.7 \leq z < 0.8$ & $(0.405,0.595)$ & $(0.326,0.674)$ & $(0.589,0.411)$ & 
     												  $(0.544,0.456)$ \\
     & $0.8 \leq z < 0.9$ & $(0.447,0.553)$ & $(0.385,0.615)$ & $(0.597,0.403)$ & 
     												  $(0.559,0.441)$ \\
     & $0.9 \leq z < 1.0$ & $(0.501,0.499)$ & $(0.471,0.529)$ & $(0.609,0.391)$ & 
     												  $(0.580,0.420)$ \\
\bottomrule
\end{tabular}
\end{table*}

Because of the simplified nature of our mocks, the \citetalias{FKP} weights are
then simply $(0.5,0.5)$ regardless of the target survey or redshift bin.
Similarly, the \citetalias{PVP} weights will be the same regardless of the
target survey or redshift bin and will result in upweighting the LRGs
as they have a higher bias -- see equation~\eqref{eq:PVPweight}. Since 
$b_{\mathrm{LRG}} \simeq 1.7b_{\mathrm{ELG}}$ at all redshifts \citep{Dawson2016,Zhao2016}, 
the \citetalias{PVP} weights are
$(0.63,0.37)$.  The PSG weights vary from redshift bin to redshift bin
somewhat, and from survey to survey, due to the varying relative number densities
of the tracers. They are also different for different model parameters. We
compute them following the numerical procedure outlined in section~\ref{sec:weighting}.

Table \ref{tab:weights} summarizes the PSG weights for the different redshift
bins of the two surveys. In some cases the PSG weights are close to the
\citetalias{FKP} or \citetalias{PVP} weights, but in many cases differ
substantially. It is also interesting to note that for the DESI mocks, the PSG
weights actually imply that above a redshift of 0.7, it would be better to
consider only the ELGs when measuring the RSD parameters. However, this does not
mean that the LRG and cross power spectra do not contain additional information.
The LRG, ELG and cross power spectra (with appropriate covariance
matrices) in principle contain all of the information. Our weighting in some
sense gives the most optimal mixture (best principle component) of the three if
they were to be reduced to a single power-spectrum, but other orthogonal
mixtures (next principle components) will of course contain additional
information. Having $w_\mathrm{LRG}=0$ simply means that a pure ELG power-spectrum is
better at constraining the $f$ parameter compared to any other mixture of ELGs and
LRGs.

\begin{figure}
\includegraphics[width=1.0\linewidth]{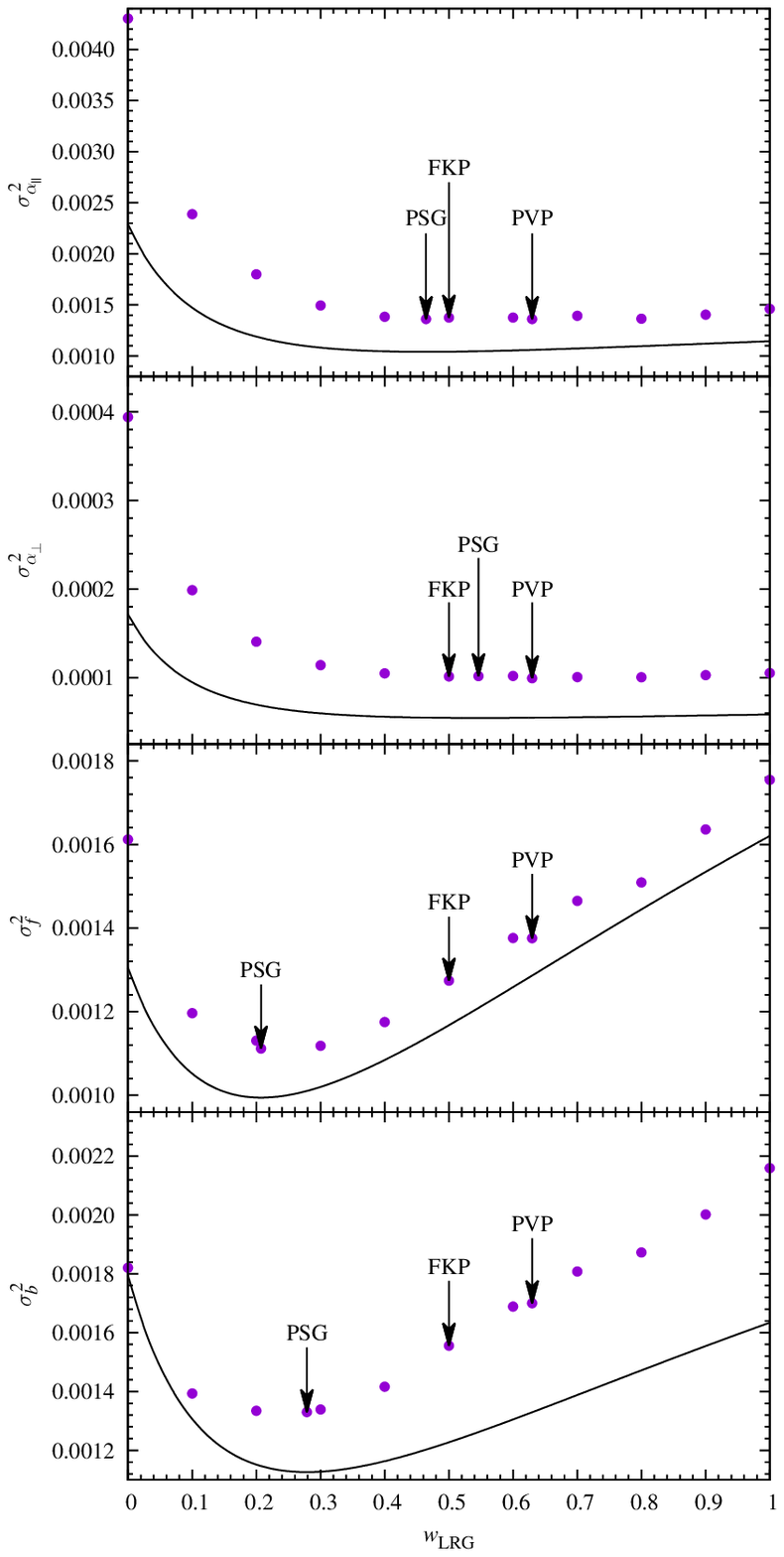}
\caption{Variance of each parameter versus the relative weight of the LRGs for the first
redshift bin of the DESI mocks. The points associated with the \citetalias{FKP}, 
\citetalias{PVP}, and PSG weights have been labelled. The
solid lines are the theoretical predictions for the variance. In general, the points follow 
the shapes of the theoretical curves. For $b$ and $f$, the PSG weights are clearly optimal,
while for $\alpha_{\perp}$ and $\alpha_{\parallel}$ the variance is flat for a broad range
of weights leading all weighting schemes to perform equally well.}
\label{fig:sigvswlrg}
\end{figure}

Fig.~\ref{fig:sigvswlrg} shows the resulting variance of the four parameters
for a variety of weights starting at $(0.0,1.0)$ --~only ELGs~-- and going to
$(1.0,0.0)$ --~only LRGs~-- in steps of 0.1, as well the \citetalias{FKP},
\citetalias{PVP} and PSG weighting schemes, in the first redshift bin for the
DESI mocks. The points show the actual variance in
measured parameter values from the mocks, while the lines show theoretical
predictions based on our Fisher matrix formalism.

It is remarkable that the measured variances follow the theoretical predictions
very closely. The fact that the minimums of the theoretical curves match well to
the measured minimum variance values shows that the methods presented in
section~\ref{sec:weighting} are sufficiently accurate and could, in principle,
be applied to any parameter that needs to be constrained, so long as non-linear
effects can be safely ignored. 

For the remaining redshift bins of the DESI mocks and all redshift bins of the
eBOSS mocks, we simply report the variance of each parameter for the \citetalias{FKP}, 
\citetalias{PVP} and PSG weights. These results are summarized in 
Table~\ref{tab:weightcomp}. It can be seen that the PSG weights derived here essentially
always produce smaller variances on their associated parameter. However, there are some
cases in which the PSG weights produce the same or larger variances than the
\citetalias{FKP} or \citetalias{PVP} weights. We find that in these cases, the theoretical
variances have a broad minimum similar to what is seen for $\alpha_{\perp}$ and
$\alpha_{\parallel}$ in Fig.~\ref{fig:sigvswlrg}. This means that we expect the gains
in those cases to be minimal at best, and small fluctuations in the measured variance
about its true value could lead to the PSG weights having a slightly higher variance 
than the other weighting schemes.

The improvements made, on the other hand, can be quite large. For example, in the second
redshift bin for the DESI mocks, the PSG weights reduce the variance in $f$ by ${\sim}36$ 
per cent compared to the \citetalias{PVP} weights. On average, the improvements for $b$ and
$f$ are ${\sim}14$ and ${\sim}10$ per cent, respectively. Yet $\alpha_\parallel$ and 
$\alpha_\perp$ seem to be insensitive to the weighting as long as the LRG weights are not 
very low.

\begin{table*}
\caption{Comparison of \citetalias{FKP} and \citetalias{PVP} weights with those
derived here (PSG). Column 1 indicates the target survey. Column 2 lists the redshift range. 
Column 3 indicates the weighting scheme. The remaining columns list the variance in
each of the parameters.}
\label{tab:weightcomp}
\centering
\begin{tabular}{@{}lcccccc@{}}
\toprule
Survey & Redshift Range & Weights & $\sigma_{b}^2$ & $\sigma_{f}^2$ & 
$\sigma_{\alpha_{\perp}}^2$ & $\sigma_{\alpha_{\parallel}}^2$ \\
\midrule
DESI & $0.6 \leq z < 0.7$ & FKP & 0.00156 & 0.00127 & 0.00010 & 0.00138 \\
     & 					  & PVP & 0.00170 & 0.00138 & 0.00010 & 0.00136 \\
     &					  & PSG & 0.00133 & 0.00111 & 0.00010 & 0.00136 \\ [1ex]
     & $0.7 \leq z < 0.8$ & FKP & 0.00078 & 0.00064 & 0.00007 & 0.00094 \\
     & 					  & PVP & 0.00093 & 0.00077 & 0.00008 & 0.00098 \\
     &					  & PSG & 0.00056 & 0.00049 & 0.00007 & 0.00097 \\ [1ex]
     & $0.8 \leq z < 0.9$ & FKP & 0.00076 & 0.00061 & 0.00007 & 0.00094 \\
     & 					  & PVP & 0.00085 & 0.00067 & 0.00007 & 0.00090 \\
     &					  & PSG & 0.00060 & 0.00052 & 0.00007 & 0.00092 \\ [1ex]
     & $0.9 \leq z < 1.0$ & FKP & 0.00060 & 0.00055 & 0.00007 & 0.00084 \\
     & 					  & PVP & 0.00064 & 0.00058 & 0.00006 & 0.00082 \\
     &					  & PSG & 0.00057 & 0.00053 & 0.00007 & 0.00085 \\ [1ex]
eBOSS& $0.6 \leq z < 0.7$ & FKP & 0.03599 & 0.02775 & 0.00343 & 0.05580 \\
     & 					  & PVP & 0.04186 & 0.03120 & 0.00320 & 0.05588 \\
     &					  & PSG & 0.03546 & 0.02709 & 0.00305 & 0.05259 \\ [1ex]
     & $0.7 \leq z < 0.8$ & FKP & 0.01867 & 0.01526 & 0.00174 & 0.02983 \\
     & 					  & PVP & 0.02100 & 0.01686 & 0.00168 & 0.02889 \\
     &					  & PSG & 0.01794 & 0.01549 & 0.00168 & 0.02862 \\ [1ex]
     & $0.8 \leq z < 0.9$ & FKP & 0.02381 & 0.02065 & 0.00257 & 0.03918 \\
     & 					  & PVP & 0.02570 & 0.02215 & 0.00245 & 0.03749 \\
     &					  & PSG & 0.02486 & 0.02213 & 0.00242 & 0.03917 \\ [1ex]
     & $0.9 \leq z < 1.0$ & FKP & 0.05472 & 0.05063 & 0.00896 & 0.09235 \\
     & 					  & PVP & 0.05980 & 0.05415 & 0.00862 & 0.08979 \\
     &					  & PSG & 0.05344 & 0.04994 & 0.00956 & 0.08834 \\

\bottomrule     
\end{tabular}
\end{table*}

\section{Conclusions}
\label{sec:conclusion}

We have presented a method of determining the relative weights that will result
in the minimal variance of cosmological parameters measured from a joint power
spectrum of multiple tracers. Tests on mock catalogues replicating eBOSS and
DESI samples show that these weights will result in a 10 to 35 per cent decrease
in the variance of the measured growth rate parameter compared to the commonly
used weighting schemes. Our weighting scheme is different from the one presented
in \citet{Hamaus2012} as it aims to find a single power spectrum (a most optimal
mixture of the tracers) that is optimal for the cosmological constraints, while
the weights in \citet{Hamaus2012} aim to split the tracers in two groups in a
way that is most optimal for the RSD parameters. The decision about which weights to
use will depend on what kind of analyses one has in mind. If the cosmological
parameters will be measured from the joint power spectrum then the PSG weights
are optimal; if the tracers will be split in two groups with a full multi-tracer
analysis to follow then the \citet{Hamaus2012} can be used to determine the most
optimal splitting.

Our derivation relies on several simplifying assumptions that are commonly
adopted when deriving optimal weights. We assume that the galaxy field is
perfectly Gaussian and calculate the variance of the power-spectrum and its 
sensitivity to cosmological parameters using linear theory.
\citet{Smith2015,Smith2016b} showed that by abandoning some of these assumptions
for the density dependent weighting the performance of the weights can be
improved by a further 20 per cent. We do not expect such a large improvement in
our case since the theoretical predictions based on our simplified treatment
eventually turn out to be very close to the actual results (see
Fig.~\ref{fig:sigvswlrg}).

The optimal weights are by definition different for different cosmological
parameters. Fortunately, it seems as though for the eBOSS and DESI samples,
related parameters have similar optimal weights. In this case  `average'
optimal weights, that are nearly optimal for all parameters of interest, can be
found. A more optimal solution would be to compute each cosmological parameter
from its own `custom weighted' power-spectrum and find the covariance between
them from the mock catalogues. Another option is to go one step beyond and find
the optimal weights for the dark energy parameters that are derived from $f$,
$\alpha_\parallel$, and $\alpha_\bot$.

A logical continuation of this work is to extend the formalism to samples
with varying number density along the redshift shell. One could also try to
incorporate it with the weights designed to optimize the measurements by
accounting for the redshift evolution of the sample. We leave these matters to a
future investigation.

\section*{Acknowledgements}
LS is grateful for the support from SNSF grant SCOPES IZ73Z0 152581, GNSF grant
FR/339/6-350/14, and NASA grant 12-EUCLID11- 0004. This work was supported in
part by DOE grant DEFG 03-99EP41093.  NASA's Astrophysics Data System
Bibliographic Service and the arXiv e-print service were used for this work.
Additionally, we wish to acknowledge \textsc{gnuplot}, a free open-source
plotting utility which was used to create all of our figures. We acknowledge the
use of the Legacy Archive for Microwave Background Data Analysis (LAMBDA), part
of the High Energy Astrophysics Science Archive Center (HEASARC). HEASARC/LAMBDA
is a service of the Astrophysics Science Division at the NASA Goddard Space
Flight Center.




\bibliographystyle{mnras}
\bibliography{2TracerWeights_v4} 



%
%


\bsp	
\label{lastpage}
\end{document}